%%% THIS IS A Plain TeX file.
%%% This is a Plain TeX file.
\magnification=1200

%%%%\baselineskip=24pt
%%\hfill {IUT-PHYS }
%%% \hfill {September  2002} \vskip 0.1in

\font\bigbf=cmbx10  scaled\magstep3 \vskip 0.2in
\centerline{\bigbf Some New Exact Results for } \vskip 0.1in
\centerline{\bigbf the q-State Potts Model on Ladder Graphs}

\vskip 0.4in \font\bigtenrm=cmr10 scaled\magstep1
\centerline{\bigtenrm  B. Mirza$^{\dag \ddag}$ and M.R.
Bakhtiari$^{\dag}$} \vskip 0.2in

\centerline{\sl $^{\dag}$Department of Physics, Isfahan University of Technology, Isfahan 84154, Iran }
\vskip 0.1in
\centerline{\sl $ ^{\ddag}$Institute for Studies in Theoretical Physics and Mathematics, }
\centerline{\sl P.O.Box 5746, Tehran, 19395, Iran}
\vskip 0.1in

%%\centerline{\sl Isfahan University of Technology,}
%%\centerline{\sl Isfahan 80971, Iran,}
\centerline{\sl E-mail: b.mirza@cc.iut.ac.ir}

\vskip 0.2in \centerline{\bf ABSTRACT} \vskip 0.1in We present
exact calculations of the partition function for the q-state
Potts model for general q, temperature and magnetic field on
strips of the square lattices of width $L_{y}=2$ and arbitrary
length $L_x = m $ with periodic longitudinal boundary conditions.
A new representation of the transfer matrix for the q-state Potts
model is introduced which can be used to calculate the determinant
of the transfer matrix for an arbirary $m \times m$ lattice with
periodic boundary conditions.

\vskip 0.1in \noindent PACS numbers: 05.50.+q, 02.70.-c

\noindent Keywords: Potts Model; Exactly Solvable Models.

\vfill\eject

\vskip 1in \centerline{I. \bf \ \  Introduction }
\vskip 0.1in

The two dimensional q-state Potts models [1,2] for various q have
been of interest as examples of different universality classes
for phase transisitions and, for q=3,4 as models for the
adsorption of gases on certain substrates [3,4,5]. For  q $\geq $3
the free energy has never been calculated in closed form for
arbitrary temperature. It is thus of continuing value to obtain
further information about the two dimensional potts model. Some
exact results have been established for the model: from a duality
relation, the critical point has been identified [1]. The free
energy and latent heat [6,7,8], and magnetization [9] have been
calculated exactly by Baxter at this critical point, establishing
that the model has a continuous, second order transition for q$
\leq $ 4 and a first order transition for q $ \geq $ 5. Baxter
has also shown that although the q $ = $ 3 model has no phase with
antiferromagnetic long-range order at any finite temperature there
is an antiferomagnetic critical point at T $ = $ 0 [9]. The values
of the critical exponents (for the range of q where the
transition is continuous) have been determined [10,11,12]. Further
insight into the critical behaviour was gained using the methods
of conformal field theory [13]. In this paper, a new
representation for transfer matrix of the q-state Potts model is
introduced. The determinant of the transfer matrices can be
calculated, which corresponds to the results suggested in [14]
for square lattices. There is a duality relation for the largest
eigenvalues of Potts model on square lattices with periodic
boundary conditions which determines a size-independent value for
the reduced internal energy at the critical point. In the last
section, an exact solution is obtained for the q-state Potts model
on ladder graphs and in a magnetic field. The paper is organized
as follows: In section II, a new representation for the transfer
matrix of the q-state Potts model is given. This representation
requires operator-like numbers. Determinant of the transfer
matrices for m $ \times $ m lattices and exact solution for the
largest eigenvalue for q-state Potts model on a 2 $ \times $ m
square lattice with periodic boundary conditions is calculated.
In section III, exact solution for the q-state Potts model in a
magnetic field and on a ladder graph is given.

\vskip 0.2in \centerline{II. \bf  \  \ New representation for the
transfer matrices } \vskip 0.1in
 The q-state Potts model has served as a valuable model in the
 study of phase transition and critical phenomena. On a lattice,
 or more generally on a graph $G$, at temperature $T$ this model is
 defined by the partition function:
$$
Z(G,q,a)=\sum_{\{\sigma_n\}} e^{-\beta\, H} \eqno(1)$$ \noindent
 with the
Hamiltonian (the critical temperature of this model is half of
the standard definition of the Potts model)
$$H=-J\sum_{<i,j>}(2\,\delta_{\sigma_i \sigma_j} - 1) \eqno(2)$$ where
$\sigma_i=1,\cdots,q$ are the spin variables on each vertex $i\in
G$ , $\beta=(k_B T)^{-1}$; and $<i,j>$ denotes pairs of adjacent
vertices. We use the notation:
$$k=\beta J\,\,;\,\,a=e^k \eqno(3)$$

\noindent
 Consider an m $\times $ m  square lattice with periodic
boundary conditions. For the Ising model, partition function can
be written as product of transfer matrices [15,16]

$$Z=Tr P^m \eqno(4)$$

$$ P = V_2 V_{1}^{\prime} \eqno(5) $$
 where
$$\eqalignno{V_2&=\prod_{\alpha=1}^{m}e^{k Z_\alpha\,
Z_{\alpha+1}} &(6) \cr
 &=\exp{(\ k\sum_{\alpha=1}^{m}Z_{\alpha}\ Z_{\alpha+1})} &(7) \cr
 V_{1}^{\prime}&=[2\sinh2k]^{{m}\over{2}}
 \exp\,(\,{\tilde{k}\sum_{\alpha=1}^{m}X_\alpha})\,\  &(8)\cr
 X_\alpha&=1\otimes\ldots\otimes1\otimes
 X\otimes1\otimes\ldots\otimes1\,\ &(9) \cr
 Z_\alpha&=1\otimes\ldots\otimes1\otimes
 Z\otimes1\otimes\ldots\otimes1\,\ &(10) \cr}$$
 \noindent
 where X and Z are $ \alpha $th factors and given by
$$ X=\left(\matrix{0&1\cr
            1&0\cr}\right)\eqno(11)$$
$$ Z=\left(\matrix{1&0\cr
            0&-1\cr}\right)\eqno(12)$$
\noindent
 and $\tilde{k} $ is given by the following duality relation (q=2):
$$e^{-2 \tilde{k}}={{e^{k}-e^{-k}}\over {e^{k}+(q-1)e^{-k}}} \eqno(13)$$
\noindent

Feynman once said that "every theoretical physicist who is any
good knows six or seven different theoretical representations for
the same physics" [17]. There are several representations for the
transfer matrix of the q-state Potts model [18,19,20] but we
introduce a new one which helps us to obtain the determinant of
the matrices easily. In this new formulation, the matrices
$V_{1}^{\prime}$
 and $X$ have to be replaced by:
 $$\eqalignno{
 V_{1}^{\prime}&=\big[2\,\sinh k \,\big(e^k +(q-1)\,e^{-k}\,)\,\big]^{m \over2}
 \,\,\exp{(\ \tilde{k}\sum_{\alpha=1}^{m}X_{\alpha})}&(14) \cr
 &=\big[2\,\sinh k \  \big(e^k+(q-1)\,e^{-k}\,)\,\big]^{m\over2}\,V_1\,\, &(15)\cr
 X&={2\over q}\,\biggl(\,\sigma+(\,1-{q\over2\,}\,)\,I\,\biggr)
 &(16) \cr}$$
 where $\sigma$ is a $q\times q$ matrix with zero diagonal
 elements and unit elements on all other entries and $I$ is a
 $q\times q$ unit matrix.
 $$
 X=\pmatrix{{2\over q}-1&{2\over q}&\ldots&{2\over q}\cr
 {2\over q}&{2\over q}-1& & & \cr
 \vdots& & \ddots& &\cr
 {2\over q}&\ldots& & {2\over q}-1}
\eqno(17) $$ $Z$ also has to be replaced by a diagonal $q\times q$
matrix and its elements satisfy the following product by
definition:
$$Z=
\pmatrix{a_1 & 0 & \ldots & 0 \cr
          0 & a_2 & \ldots & 0 \cr
          \vdots & \vdots & \ddots & \vdots  \cr
          0 & 0 & \ldots & a_q \cr}\eqno(18)$$
 $$a_i\,a_j=2\,\delta_{i j}-1 \eqno(19)$$

\noindent
   Expect $q=2$ case, one cannot represent $a_i$ elements by real
   or complex numbers because of the defined product (19); however, elements
   of $V_2$ and the transfer matrix are real numbers(in calculation of $V_2$ only product of
   two $a_i $ elements appear). The determinant of the transfer matrix can be calculated easily.
   $$
   \eqalignno{\det V_1&=\exp\,\bigg[Tr
   \big(\tilde{k}\sum_{\alpha=1}^{m}\,X_\alpha\big)\bigg] &(20)\cr
    &=\exp\bigg[m\,\tilde{k}\,q^{m-1}\,(2-q) \bigg]
   &(21) \cr}$$
and
$$\eqalignno{\det V_2&=\exp\,\bigg[ Tr
   \big({k}\sum_{\alpha=1}^{m}\,Z_\alpha
   Z_{\alpha+1}\big)\bigg] &(22) \cr
    &=\exp\bigg[m\,k\,q^{m-2}\,\sum_{i,j}a_i\,a_j \bigg]&(23)\cr
     &=\exp\bigg[m\,k\,q^{m-1}\,(2-q) \bigg]&(24)\cr
}$$

\noindent
and

$$ \det(V_2 V_1 )=\exp [-m {q^{m-1}} (q-2)(k + \tilde{k}) ]\eqno(25) $$

\noindent
 $ \det(V_2 V_1 ) $ has a maximum at the self dual point
$k=\tilde{k}$ and finally determinant of the transfer matrix is
 given by,

 $$\det P =[2 \ (e^{k}+(q-1)e^{-k})\  \sinh k ]^{{m^2}\over 2} \  \exp [-m {q^{m-1}} (q-2)(k + \tilde{k})] \eqno(26)$$
\noindent
 For a $2\times m $ square lattice one can calculate the
largest eigenvalue of the transfer matrix using different methods
[21, 22]. For the Hamiltonian (2) the largest eigenvalue for the
transfer matrix is given by  ,

$$ \lambda_{max}(k,\tilde{k})=[2 \ (e^{k}+(q-1)e^{-k})\ \sinh k \ ]\ \exp[\gamma ]\eqno(27)$$

\noindent
where $\gamma $ is given by

$$\cosh \gamma = \cosh[2 k]\cosh[2 \tilde k] - ({{q-2}\over q}) \  \sinh[2k]\sinh[2 \tilde k] \eqno(28)$$

\noindent
 The above solution for q=2 is equal to the exact
solution for a $2 \times m$ square lattice obtained by Onsager
[23]. For an arbitrary $m \times m $ lattice, the largest
eigenvalue for q state Potts model has a general form which is the
result of duality, for duality in Potts model see[18,24].

$$ \lambda_{max}(k,\tilde{k})=[2 \ (e^{k}+(q-1)e^{-k})\ \sinh k \ ]^{m \over 2} \ f(k,\tilde k)\eqno(29) $$
where $ f(k,\tilde k) $ is a symmetric function of $k $ and $
\tilde k $ and is not known ($q \geq 3$ ) for an arbitrary m
$\times $ m lattice.

\noindent
 Using the same method, one can
prove the formula for the determinant of the transfer matrix
which has been conjectured in reference [14]. The Hamiltonian is
defined by
$$H=-J\sum_{<i,j>} \delta_{\sigma_i \sigma_j} \eqno(30)$$
\noindent and the matrices $X $  and  $ V_{1}^{\prime} $  and the
duality relation are given by
$$e^{- \tilde{k}}={{e^{k}-1}\over {e^{k}+(q-1)}} \eqno(31)$$
 $$\eqalignno{
 V_{1}^{\prime}&=\big[e^k -1 \big]^{m} \,\,\exp{(\ \tilde{k}\sum_{\alpha=1}^{m}X_{\alpha})}&(32)\cr
 &=\big[e^k-1 \big]^{m }\,V_1\,\,;&(33)\cr
 X&={2\over q}\,\biggl(\,\sigma+(\,1-{q\over2\,}\,)\,I\,\biggr)&(34) \cr}$$
and $V_2 $ is given by the same formula as before. The definition
of the product of the elements of the $Z $ matrix is different
and is defined by
 $$a_i\,a_j=\delta_{i j} \eqno(35)$$
 After straightforward calculation the determinant of the transfer
 matrix in this case is given by

$$\det P =(e^{k}-1)^{m\, q^m} (e^k \,  e^{\tilde{k}})^{m \,q^{m-1}}\eqno(36) $$
It may be interesting to extend these results to lattices with
different boundary conditions and to prove all the determinants
which are conjectured in [14].

\vskip 0.2in \centerline{II. \bf  \  \ Potts model on ladder
graphs and in a magnetic field } \vskip 0.1in

Exact solution of the two dimensional Ising model in a magnetic
field is not known; however, on a one dimensional lattice 1 $
\times $ m the partition function can be calculated (e.g. see
[25]). The simplest generalization is to solve this problem for
 ladder graphs (2 $\times$ m ) with longitudinal periodic boundary
condition. There could be several applications for this solution.
On a ladder graph the transfer matrix for the q state Potts model
is of order
 $2^q $. For $q=2 $, the transfer matrix is of order 4 and the characteristic equation
can be obtained. It is posible to do the same thing for 3, 4, 5,
and 6 state Potts model and then generalize it to an arbitrary q.
Consider the following Hamiltonian for the q state Potts model on
a ladder graph with longitudinal periodic boundary conditions.

$$ H=\sum_{<i,j>}(-J \, \delta_{\sigma_i , \sigma_j}- h \, \delta_{\sigma_i
,1}) \eqno(37)$$

\noindent where h is the magnetic field. The partition function is
given by

$$ Z(G,q,x,y)=\sum_{\{\sigma_n\}} e^{-\beta\, H} \eqno(38) $$
\noindent where
$$ x= e^{\beta J}; \, y=e^{\beta h} \eqno(39)$$

\noindent Our result for the partition function of the q state
Potts model on a 2 $\times $ m lattice with longitudinal periodic
boundary condition is given by,

$$ Z=\sum_{j=1}^4 \, \lambda_{4,j}^m \,+ \,  (q-2)\sum_{j=1}^3 (\lambda_{3,j}^m \, + \, \lambda_{2,j}^m) \,+
 \, (q^2-5q +5)\, \lambda_{1,1}^m \, +\,{\lambda_{1,1}^{\prime}}^m \eqno(40)$$

\noindent where $ \lambda_{i,j} $ is the jth root of an equation
of order i. $ {\lambda_{1,1}^{\prime}}$ is a root for the
following first order equation,
$$ \lambda + (q-1)\, y +(2-q)\, xy -x^2 y=0 \eqno(41)$$

\noindent  $ {\lambda_{1,1}}$ is a root for the following
equation,

$$( \lambda - x^2 + 2 x -1)^{q^2 -5q +5}=0 \eqno(42)$$

\noindent The terms $\lambda_{2,j} $ for $j=1,2$ are the roots of
the second order equation,

$$ [ \lambda^{2} +  c_{21}\,  \lambda  + c_{22}]^{q-2} =0 \eqno(43)$$

\noindent with
$$c_{21} = (q-2)+(3-q)x-x^2+xy-x^2y \eqno(44)$$

$$ c_{22}=(1-q)y +(3q-4)xy-(3q-6)x^2 y +(q-4)x^3 y + x^4 y  \eqno(45)$$

\noindent The terms $\lambda_{3,j} $ for $j=1,2,3$ are  roots of
the cubic equation,
$$(\lambda^{3}+ \lambda^{2} \, c_{31}+\lambda \, c_{32} + c_{33})^{q-2}=0 \eqno(46)$$
with
$$ c_{31}=q-4 +(6-q)x-x^2 -x^3+xy-x^2 y \eqno(47)$$
$$ c_{32}=(2-q)x+(3q-7)x^2-(3q-9)x^3+(q-5)x^4+x^5 +$$
$$ (3-q)y+(17-3q)x^2 y +(q-9)x^3 y +x^5 y+(3q-12)xy \eqno(48)$$
$$c_{33}=-xy(x-1)^5(x+q-1) \eqno(49)$$
\noindent The terms $\lambda_{4,j} $ for $j=1,2,3,4$ are  roots of
an algebraic equation of degree four,

$$\lambda^4 + c_{41} \lambda^3+ c_{42} \lambda^2 +
c_{43} \lambda + c_{44} =0 \eqno(50)$$

\noindent with
$$ c_{41}=-x^3 y^2 -x^3-x^2-(q^2-5q+7)-(3q-8)x-(x^2+(q-2)x+(q-1))y \eqno(51)$$
$$ c_{42}=(x-1)(q+x-1)x^3 y^3+(x-1)(x^4+2x^3+(3q-6)x^2+(q-1)^2
x+(q-2)(q-1))xy^2+$$
$$(q-5)(q-4)(q-3)(x-1)y+(x-1)(x^4+q x^3+(6q-13)x^2+(4q^2-18q+21)x+$$
$$(6q^2-35q+51))y+(q+x-2)^2(x-1)^2 x \eqno(52)$$
$$ c_{43}=-(x(x-1)^3((q+x-1)(x^3+3x^2+(3q-5)x+(q-2)(q-1))y^3+ $$
$$ (q+x-1)(x^2+(q-2)x+(q-1))x y^2+(q+x-2)^2(q+x-1)y))\eqno(53)$$
$$ c_{44}=(x-1)^5(q+x-1)^3 x^2 y^3 \eqno(54)$$

\noindent This solution is interesting as one may use it to
generalize the results of [26,27] and to identify exact location
of the partition function zeros, namely Yang-Lee[28] ,
Fisher(complex temperature)[29] and Potts (complex q) zeros for
the Potts model on the ladder graphs. Another application could
be a generalization of the arguments given in [30,31] from one
dimensional lattices to the ladder graphs.

\vskip 0.4in \centerline{V. \bf \ \  Conclusion} \vskip 0.1in
\noindent In this work a new representation for the transfer
matrix of the q state Potts model on an square lattice is
introduced. Exact result for the partition function of the q state
Potts model for general q, temperature and magnetic field on
strips of the square lattices 2 $\times $ m with periodic boundary
conditions are given. It will be interesting to generalize some
exact results for the one dimensional lattices [27,30,31] to the
ladder graphs.

\vskip 0.2in
\centerline{\bf \ \ Acknwoledgements}
\vskip 0.2in
Our thanks go to the Isfahan University of Technology and Institute for Studies in
 Theoretical Physics and Mathematics for the financial support they made available to us.
\vskip 0.2in
\centerline{\bf \ \  References}
\vskip 0.1in

\noindent [1] \ R. B. Potts, Proc. Camb. Phil. Soc. {\bf 48}, 106
(1952).

\noindent [2] \ F. Y. Wu, Rev. Mod. Phys. {\bf 54} (1982) 235.

\noindent [3] \ S. Alexander, Phys. Lett. {\bf A54}, 353 (1975).

\noindent [4] \ A. N. Berker, S. Oslund, and F.Putnam, Phys. Rev.
{\bf B17}, 3650 (1978).

\noindent [5] \ E. Domany et al, Phys. Rev. {\bf B18}, 2209
(1978).

\noindent [6] \ R. J. Baxter, J. Phys. C {\bf 6} L445 (1973).

\noindent [7] \ R. J. Baxter et al, Proc. Roy. Soc. London, Ser.
A {\bf 358}, 535 (1978).

\noindent [8] \ R. J. Baxter, J. Stat. Phys. {\bf 28},1 (1982).

\noindent [9] \ R. J. Baxter, Proc. Roy. Soc. London, Ser. A {\bf
383}, 43 (1982).

\noindent [10] \ M. P. M. den Nijis, J. Phys A {\bf 12}, 1825
(1979); Phys. Rev. {\bf B27}, 1674.

\noindent [11] \ J. L. Black and V. J. Emery, Phys. Rev. {\bf
B23}, 429 (1981).

\noindent [12] \ B. Nienhuis, J. Appl. Phys. {\bf 15}, 199 (1982).

\noindent [13] \ V. S. Dotsenko, Nucl. Phys. {\bf B235}, 54
(1984), and refs therein.

\noindent [14] \ S. Chang and R. Schrok, Physica  A {\bf  296},
234-288 (2001). cond-mat/0011503.

\noindent \ \ \ \ \ \ (e.g. Eq. 3.4.39)

\noindent [15] \ K. Huang, Statistical Physics, 2nd edition,
(Wiley). Chapter 15.

\noindent [16] \ B. Bergersen and M. Plischke, Equilibrium
Statistical Physics, 2nd edition,

\noindent \ \ \ \ \ \ (World Scientific). Chapter 5.

\noindent [17] \ R. Feynman, The character of Physical Law (MIT
Press, 1965), p. 168.

\noindent [18] \ L. Mittag and M. J. Stephen. J. Math. Phys. Vol.
{\bf 12}, No. 3 (1971).

\noindent [19] \ R. J. Baxter, Exactly Solved Models in St. Phys.
(Academic Press, 1982).

\noindent [20] \ P. Martin, Potts Models and related Problems in
St. Phys. (World Scientific).

\noindent [21] \ R. Shrock, Physica A {\bf 283},  388-446 (2000);
cond-mat/0001389.

\noindent [22] \ M. A. Yurishchev, cond-mat/0111401.

\noindent [23] \  L. Onsager, Phys. Rev. {\bf 65}, 117-49 (1944).

\noindent [24] \ F. Y. Wu and Y. K. Wang, J. Math. Phys. {\bf 17},
439 (1976).

\noindent [25] \ F. Y. Wu, cond-mat/9805301.

\noindent [26] \ Z. Glumac  and K. Uzelac, J. Phys A: Math. Gen.
{\bf 27} 7709-17 (1994).

\noindent [27] \ R.G. Ghulghazaryan and N. S. Ananikian,
cond-mat/0204424.

\noindent [28] \ C. N. Yang and T. D. Lee, Phys. Rev. {\bf 87}
404-410, (1952).

\noindent [29] \ M. E. Fisher, Lectures in Theoretical Physics,
Vol 7 C, ed W. E. Brittin

\noindent \ \ \ \ \ \ (Boulder, Co: University  of Colorado Press)
pp 1.

\noindent [30] \ P.B. Dolan and D. Johanston, 1D Potts, Yang-Lee
Edges and Chaos; cond-mat/0010372

\noindent [31] \ B. P. Dolan, D. A. Johanston, R. Kenna, The
Information Geometry of the

\noindent \ \ \ \ \ \ One Dimensional Potts Model;
cond-mat/0207180

\vfill\eject

\bye